\documentstyle[11pt]{article}
 
\def\pa{\partial}

 \def\G{\Gamma}

\def\l{\lambda} 
\def\m{\mu}

\def\x{\chi}
\def\p{\pi}

\renewcommand{\baselinestretch}{2.0}

\begin{document}

\begin{flushright}
\begin{tabular}{l}
       BRX--TH--333  \\
       hep-th/9206045
\end{tabular}
\end{flushright}

\vspace{1in}

\begin{center}
{\Large\bf Concerning the Double Scaling Limit in the O(N) Vector\\
Model in Four-Dimensions}\footnote{Supported in part by the DOE
under grant DE--FG02--92ER40706.}

by\\

\vspace{.2in}

\renewcommand{\baselinestretch}{1}
\small
\normalsize
Howard J. Schnitzer\\
Department of Physics\\
Brandeis University\\
Waltham, MA 02254

\renewcommand{\baselinestretch}{2}
\small
\normalsize

May 1992

\end{center}

\vspace{.3in}

{\bf Abstract}

\begin{quotation}
The 1/N expansion for the O(N) vector model in four dimensions is
reconsidered.  It is emphasized that the effective potential for this
model becomes everywhere complex just at the critical point, which
signals an unstable vacuum.  Thus a critical O(N) vector model cannot
be consistently defined in the 1/N expansion for four-dimensions, which
makes the existence of a double-scaling limit for this theory doubtful.
\end{quotation}

\newpage

The matrix model approach \cite{001,002} to $c \leq 1$ matter coupled
to two-dimensional gravity has provided a number of new insights, as
well as a non-perturbative definition of the sum over all genus in string
theory.  An essential feature of these methods is the existence of a
critical value of a coupling constant $g_c$, and a double scaling limit
$N \rightarrow \infty$ and $g \rightarrow g_c$, such that an
appropriate correlated scaling variable is kept fixed.

It is an interesting question to examine other field theories which allow
for a double scaling limit.  In particular the O(N) vector model
\cite{003,004,005} and generalized vector models \cite{006} have been
studied for dimensions D=0,1,2, and 3 in the double scaling limit
\cite{007,008,009}.  The
possibility of a double scaling limit for the
$(\Phi^2)^2$ O(N) vector model has also been considered for D=4
\cite{008,009}, which is the subject of this note.  We will argue that
precisely at the critical point the effective potential for the D=4 theory
becomes everywhere complex, which undermines the consistency of the
D=4 O(N) vector model at the critical point, and thus makes the
existence of a double scaling limit doubtful.  This problem does not arise
for D$<$4.

The 1/N expansion of the O(N) symmetric $\l\Phi^4$ theory in D=4
has been studied extensively \cite{003,004,005}.  We follow the analysis
of ref. \cite{004}.  The unrenormalized Lagrangian is defined by
\begin{equation}
{\cal L} = {\textstyle{\frac{1}{2}}} (\pa_\m\Phi )^2 -
{\textstyle{\frac{1}{2}}} \m^2_0 \; \Phi^2 -
\frac{\l_0}{4!N} \; (\Phi^2)^2
\end{equation}
where $\Phi_a$ is an N-component quantum field, and
$\Phi^2 = \sum^N_{a=1} \; \Phi_a \; \Phi_a$.  To leading order in
1/N, the effective potential satisfies
\begin{equation}
\frac{dV(\phi^2)}{d\phi^2} = \textstyle{\frac{1}{2}} \; \x (\phi^2)
\end{equation}
where $\x$ is related to the constant classical field $\phi$ by the gap
equation
\begin{equation}
\x = \m^2_0 + \frac{\l_0}{6} \; \left(\frac{\phi^2}{N}\right) +
\frac{\l_0}{6} \; \int \; \frac{d^4k}{(2\p )^4} \;
\frac{1}{k^2 + \x}
\end{equation}
where the integral is over Euclidean momenta.  One introduces
renormalized parameters $\m^2, \; \l$, and $g$, as well as
renormalization group invariant quantities $\x_{_0}$ and $\rho
(\phi^2)$,
with
\begin{equation}
\x (\phi^2) = \rho (\phi^2) \, \x_{_0} \;\; .
\end{equation}
The gap equation becomes
\begin{equation}
\rho \ln \rho = -\frac{96\p^2}{\x_{_0}} \; \left(\frac{\m^2}{g} \right)
-
\frac{16\p^2}{\x_{_0}} \; \left(\frac{\phi^2}{N}\right)
\end{equation}
which shows that $\x (\phi^2)$, and hence $V(\phi^2)$, is determined
once the two renormalization-group invariant parameters $\x_{_0}$ and
$(\m^2 /g)$ are specified.

{}From Figure 1 of ref 4, one observes that $\rho (\phi^2)$ is a
double-valued function of $\rho \ln \rho$ for $-e^{-1} \leq \rho \ln
\rho \leq 0$.  These two branches are defined by
\begin{eqnarray}
\mbox{\rm branch I:}   & \rho_{\rm I} < e^{-1}  \nonumber\\
\mbox{\rm branch II:}   & \rho_{\rm II} > e^{-1}
\end{eqnarray}
and where branches I and II of $V(\phi^2)$ denote that portion of the
effective potential determined by the corresponding two branches of
$\rho (\phi^2)$.  Thus the effective potential $V(\phi^2)$ has two
branches for $\phi^2 < \phi^2_b$, and was shown to be everywhere
complex for $\phi^2 \geq \phi^2_b$, where $\phi_b$ denotes the
branch point of the effective potential in the notation of ref. \cite{004}.
Further
\begin{equation}
Re \; \frac{dV(\phi^2)}{d\phi^2} \; _{\stackrel{\textstyle \sim}{\phi^2
\rightarrow \infty}} \;
\frac{-8\p^2 (\phi^2/N)}{\ln (\phi^2)}
\end{equation}
so that the potential has no lower bound.
[See Figures 2, 3, and 4 of ref. \cite{004}].  Branch I of $V(\phi^2)$
always lies above branch II, and Green's functions built on vacuua
defined by branch I always contain tachyons.  On the other hand
Green's functions constructed with respect to a vacuum defined by
branch II are free of tachyons in each order of the 1/N expansion.  This
requires $\rho_{\rm II} > e^{-1}$, or equivalently $\phi^2 < \phi^2_b$.
[However, even though a consistent perturbative 1/N expansion can be
defined with respect to vacuua of branch II, the theory is unstable to
(non-perturbative) tunneling to $\phi^2 \rightarrow \infty$].

Consider the inverse propagator for the composite field $\x$, computed
for a vacuum defined by branch II.  The result is \cite{004}
\begin{equation}
N^{-1} \, D_{\x\x}^{-1} (p) \; = \; \frac{1}{32\p^2}\; [1 + \ln
\rho_{\rm II}(\phi^2)] \; - \; 3[f(0,m^2) - f(-p^2,m^2)]
\end{equation}
where $m = \x (\phi^2)$ and
\begin{equation}
f(-p^2,m^2) \; = \; \frac{1}{48\p^2} \;
\sqrt{\frac{p^2 + 4m^2}{p^2}} \; \ln
\left[\frac{\sqrt{p^2} + \sqrt{p^2 + 4m^2}}{2m} \right] \;\; {\rm for}
\;\; p^2 \geq 0 \; .
\end{equation}
with the propagator defined by analytic continuation for other values
of $p^2$.  This is identical to the 2--point function defined by
DiVecchia, {\it et al.}, \cite{009}, {\it i.e.},
\begin{equation}
- \G (p) = N^{-1} \; D_{\x\x}^{-1} (p)
\end{equation}
where $\G (p)$ is equation (3.45) of ref. \cite{009}.  In particular our
renormalization group invariant parameter
\begin{equation}
-[1 + \ln \rho_{\rm II}(\phi^2)] = \frac{4\p^2}{f(\hat{\mu})} + \ln
\left( \frac{\hat{\mu}^2}{m^2} \right)
\end{equation}
where the right-hand side of (11) is the parameterization of ref.
\cite{009}.

The critical point of the theory is defined by $\G (0) = 0$, which
requires
$$
1 + \ln \rho_{\rm II}(\phi^2) = 0 \eqno{(12a)}
$$
or equivalently
$$
\rho_{\rm II}(\phi^2) = e^{-1} \eqno{(12b)}
$$
which corresponds to the bound-state mass $m^2_B = 0$, and
\renewcommand{\theequation}{\arabic{equation}}
\setcounter{equation}{12}
\begin{equation}
(\phi^2_b /N) = 0
\end{equation}
[c.f. Equation (5.25b) and (5.29) of ref. \cite{004}].  Equation (13)
implies from the analysis of ref. \cite{004} that the leading term in the
1/N expansion of the effective potential $V(\phi^2)$ becomes
everywhere complex, just at the critical point defined by $m^2_B = 0$.
The next to leading term in the 1/N expansion for $V(\phi^2)$ is also
everywhere complex when $m^2_B = 0$ \cite{004,005}.  Presumably
this is a feature of every order in 1/N.

Since an effective potential that
is everywhere complex signals an unstable vacuum, we conclude that
a critical O(N) vector model
cannot be consistently defined in the 1/N expansion for D=4, and
therefore there the double scaling limit for this theory is problematical.

\end{document}